\documentclass[useAMS,usenatbib]{aa}
\usepackage{txfonts}
\usepackage{natbib}
\bibpunct{(}{)}{;}{a}{}{,}
\usepackage{longtable}
\usepackage[dvips]{graphicx}
\begin{document}
\title{Probing the evolving massive star population in Orion with kinematic and radioactive tracers}
\author{R. Voss\inst{1,2} \and R. Diehl\inst{1} \and J. S. Vink\inst{3} \and D. H. Hartmann\inst{4}}
\institute{Max-Planck-Institut f\"ur extraterrestrische Physik, 
Giessenbachstrasse, D-85748, Garching, Germany
\and Excellence Cluster Universe, Technische Universit\"at M\"unchen, Boltzmannstr.
2, D-85748, Garching, Germany
\and Armagh Observatory, College Hill, Armagh, BT61 9DG, Northern Ireland, UK
\and Department of Physics and Astronomy, Clemson University, Kinard Lab of Physics, Clemson, SC 29634-0978
\\
email:rvoss@mpe.mpg.de}
\titlerunning{The evolving massive star population in Orion}
\date{Received ... / Accepted ...}

\offprints{R. Voss}

\abstract
{
Orion is the nearest star-forming region that hosts a significant 
number of young and massive stars. The energy injected by these OB stars 
is thought to have created the Eridanus superbubble. Because of its proximity, 
Orion is a prime target for a detailed investigation of 
the interaction between massive stars and their environment. 
}
{
We study the massive star population of Orion and its feedback in terms of energy and mass, in order to compare 
the current knowledge of massive stars with kinematic and radioactive tracers
in the surrounding interstellar medium (ISM).
}
{
We assemble a census of the most massive stars in Orion,
then use stellar isochrones to estimate their masses and ages, and use 
these results to establish the stellar content of Orion's individual
OB associations. From this, our new population synthesis code is utilized to derive 
the history of the emission of UV radiation and kinetic energy of the material 
ejected by the massive stars, and also follow the ejection of the long-lived
radioactive isotopes $^{26}$Al and $^{60}$Fe. 
In order to estimate the precision of our method, we compare and contrast 
three distinct representations of the massive stars. We compare the expected
outputs with observations of  $^{26}$Al gamma-ray signal and the extent of the
Eridanus cavity.
}
{
We find an integrated kinetic energy emitted by the massive stars
of 1.8$^{+1.5}_{-0.4}\times10^{52}$ erg. 
This number is consistent with the energy thought to be required to
create the Eridanus superbubble. We also find good agreement between
our model and the observed $^{26}$Al signal, estimating a mass of 
$5.8^{+2.7}_{-2.5}\times10^{-4}$ $M_{\odot}$ of $^{26}$Al in the 
Orion region. 
}
{
Our population synthesis approach is demonstrated for the 
Orion region to reproduce three different kinds of observable outputs 
from massive stars in a consistent manner: Kinetic 
energy as manifested in ISM excavation, ionization as manifested in 
free-free emission, and nucleosynthesis ejecta as manifested in 
radioactivity gamma-rays. The good match between our model and the
observables does not argue for considerable modifications of 
mass loss. If clumping effects turn out to be strong, other processes 
would need to be identified to compensate for their impact on massive-star 
outputs. Our population synthesis analysis jointly treats kinematic output
and the return of radioactive isotopes, which proves a powerful extension 
of the methodology that constrains feedback from massive stars.
}


\maketitle

\section{Introduction}
At a distance of only some 400 pc, the Orion region is sufficiently 
close that it enables us to study its stellar population 
and interstellar-gas morphology in detail. 
The massive star ($M$ $>$ 8$M_{\odot}$) population  is dominated by the Orion OB1
association, which includes four subgroups labelled a-d \citep{Brown1994}. 
Their ages have been estimated to range between 1 and 12 Myr. OB1 is located on the near side of 
the densest part of the Orion Molecular clouds \citep{Maddalena1986}, 
facing the Eridanus cavity which extends from these molecular clouds towards the Sun. H$\alpha$ features, 
which coincide with a hole in the HI distribution 
\citep{Heiles1976}, together with X-ray emission near HI features 
\citep{Burrows1993} outline this large interstellar cavity, and 
provide evidence for the interactions between the hot gas in the Eridanus 
cavity and the neutral surrounding interstellar medium \citep{Bally1991}.
Many different names are used for the various
substructures of Orion. We use the same convention as 
\citet{Bally2008} in his recent review.

Feedback from massive stars plays a crucial role in the formation
of stars, as it is shaping the ISM and its subsequent star formation activity. 
The main feedback
originates from the ejection of matter from massive stars through their winds
and supernova explosions, and from their intense emission at short wavelengths into the UV. 
This UV radiation creates large photoionized regions around the stars, and 
the kinetic energy associated with ejection of stellar matter pushes at the ISM, 
together creating large shells and cavities \citep[e.g.][]{Heiles1976,vanderHucht1987,Leitherer1992,Maeder1994}. 
Kinetic energy output and UV radiation of massive stars were 
studied in \citet{Voss-popsyn}, discussing the total emission from a
population of stars with emphasis on the differences
between various alternative stellar models. It was found
that the kinetic energy from winds dominates over the supernova 
contribution, when integrated over the first 10 Myr after the stars were formed
 \citep[see also][]{Leitherer1999}. This is due to the high wind velocities  %% do we only want to credit this to Leitherer?
\citep[taken from][]{Howarth,Lamers1995} and high mass-loss rates 
of the most massive stars \citep{Castor1975} and \citep{Vink2000}, even 
when modest wind clumping (with clumping factors of about 5) 
is accounted for \citep{Repolust2004,Mokiem2007}. To understand the complex
interplay between the massive stars and their local environment,
which eventually result in the evolution of disks in galaxies, it is necessary
to first create a census of the radiation, energy, and matter output of individual
nearby star-forming regions like Orion, which can be validated in terms of observational constraints.

The radioactive isotope $^{26}$Al provides an interesting independent view
on the interaction between young stars and the surrounding
environment. It is traced by its $\gamma$-ray decay line at
1808.63 keV, which can be observed with $\gamma$-ray telescopes. 
With a mean lifetime of $\sim$1 Myr $^{26}$Al is a long-term tracer
of nucleosynthesis from populations of massive-star sources, as they eject it
after synthesis in stellar cores and the supernova itself \citep{Prantzos1996}.
Typically, massive stars eject a few 
$10^{-5}M_{\odot}$ of $^{26}$Al through their winds and supernovae (SN) 
 \citep[see e.g.][]{Limongi2006}. From $\gamma$-ray observations,
the total mass of $^{26}$Al in the Milky Way is estimated to be
2.8$\pm$0.8$M_{\odot}$ \citep{Diehl2006}.
Measurements of the $^{26}$Al emission from Orion by the 
COMPTEL instrument on NASAs Compton observatory generally confirms this scenario,
with a $\gamma$-ray intensity of $\sim$ 7.5~10$^{-5}$~ph~cm$^{-2}$ s$^{-1}$ \citep{Diehl2002}.
The map of this $^{26}$Al $\gamma$-ray emission, though not significant in its details, 
shows an interesting offset of the $^{26}$Al emission
from the massive stars that are believed to the the source of the $^{26}$Al,
and the emission appears rather extended \citep{Diehl2002}.
This suggests that the radioactive
ejecta stream  into the nearby Eridanus cavity from their stellar association sources.
A similar tracer would be the isotope $^{60}$Fe, observed in the Galaxy globally by its 1173 keV and
1333 keV decay lines \citep{Smith2005,Harris2005,Wang2007}. This isotope is presumably created in neutron capture reactions
in late shell burning stages of such massive stars, and is
also emitted in the supernova explosions \citep{Limongi2006}.
$^{60}$Fe has a mean lifetime of $\sim$3.6 Myr \citep{Rugel2009}. It has not been seen from the Orion region,
which is however not surprising, as its $\gamma$-ray intensity has been found to be $\sim$15\% of the $^{26}$Al $\gamma$-ray intensity only.

In this paper we analyze
the energy and radioactive-isotope output from the entire stellar content of the Orion region.
We compare the results with observational constraints, such as the measured
strength of the 1808.63 keV line from $^{26}$Al decay and the size of the Eridanus
superbubble.

%%%%%%%%%%%%%%%%%%%%%%%%%%%%%%%%%%%%%%%%%
\section{The massive-star content of Orion}
Star formation in the Orion region is distributed over a number of
distinct groups. We concentrate our analysis onto the 5 most
massive groups, which are the 4 subgroups of the Orion OB1 association
\citep{Blaauw1964}:
\begin{itemize}
\item {\tt OB1a} is located to the northwest of Orion's Belt region. 
\item {\tt OB1b} defines the belt region itself. It contains three O stars,
$\zeta$ Ori and $\delta$ Ori, which together with the B star
$\epsilon$ Ori form Orion's belt, and $\sigma$ Ori as another prominent member.
\item {\tt OB1c} partially overlaps with OB1b, extending from the
Belt to the end of Orion's Sword. One O star, $\iota$ Ori, the brightest star
in the Sword, also belongs to this group.
\item {\tt OB1d} is also called the Orion Nebular Cluster. 
It contains two O stars, $\theta^1$ Ori (Trapezium) and $\theta^2$ Ori.
\item {\tt $\lambda$ Ori}. This group is often not included in 
the lists of associations with OB stars in Orion. 
However, it is located near the OB1 association, at
the Head of Orion, and at approximately the same distance. We
therefore include it in our study. It contains a single O star,
$\lambda$ Ori, after which the association is named.
\end{itemize}
From published data we investigate the properties of each of these
5 groups individually. The results constitute assumptions used in our analysis, and are
summarized in table \ref{tab:regs}.

The inventory of massive stars above 2 $M_{\odot}$
in the OB1 associations was analyzed in \citet{Brown1994}, who estimated
a total of $\sim$610 stars in the four groups, with similar numbers 
$\lesssim$200 in OB1a,b and c. In their table 4, they list the numbers
of stars found and the mass ranges probed in the three groups. We
combine these with the Salpeter mass function to estimate the total
initial number of stars in each group. This gives a total of 420 stars,
only $\sim2/3$ of the result of \citet{Brown1994}, who used a much
steeper $\alpha=2.7$ initial mass function. OB1d was found to host 
145 stars more 
massive than 1 $M_{\odot}$ \citep{Hillenbrand1997}, making it the 
smallest of the 4 OB1 subgroups, and $\lambda$ Ori contains
$\lesssim50$ stars above a mass of 2.5$M_{\odot}$ \citep{Dolan2001}.
For comparison we convert these numbers
into the 2-120 $M_{\odot}$ range using the \citet{Salpeter1955} mass function.

The ages and distances of the individual clusters vary somewhat between
publications, and are not yet agreed on in the community
\citep[see e.g. the recent reviews of][]{Bally2008,Muench2008,Walter2008,
Mathieu2008,Briceno2008}. However, most agree that the 4 associations
form a sequence in age and distance, with OB1a being the oldest and
nearest and OB1d the youngest and most distant \citep[although in the
study of][OB1b is significantly younger than OB1c]{Brown1994}. While
there are large uncertainties in the absolute distances to the groups,
the relative distances are much better understood, and it is therefore
very unlikely that they are all at the same distance. This is due to
the fact that the systematics affecting the distances are the same
for the four regions. For a compilation and thorough discussion of 
the distance studies, see \citet{Muench2008}.
The picture is complicated by the partial overlap of the groups and the
possibility that they themselves consist of several distinct subgroups
with different ages and distances \citep[see e.g.][]{Hardie1964,Warren1977,Guetter1981,Gieseking1983,Genzel1989}. 
In the following work we adopt a 
distance of $\sim$ 410 pc to OB1d as an average of the
three recent determinations of \citet{Hirota2007,Sandstrom2007,Menten2007},
see e.g. the review of \citet{Muench2008}, which is also consistent with
the results of \citet{Jeffries2007}. 
OB1c is slightly closer than the
OB1d group, which places it at a distance of $\sim$ 400 pc \citep{Muench2008}.
For OB1b we adopt a distance of $\sim$360 pc \citep{Brown1994}, and for
OB1a a distance of $\sim$330 pc \citep{Briceno2005,Briceno2007}. We note
that other relatively recent determinations find distances almost 100 pc 
further away for some of the subgroups (see discussions in the reviews 
mentioned above), and that disagreement between different methods is
significantly above the typical $\sim$10\% errors. We adopt a distance of 
$\sim$450 pc to $\lambda$ Ori \citep{Dolan2001}.

It is clear that Orion OB1a is the oldest of the groups with an age
of $\sim$8-12 Myr \citep{Blaauw1964,Warren1977,Brown1994,Briceno2005}.
OB1d, where star formation is still underway, is clearly the
youngest group consisting of stars with ages 0-2 Myr 
\citep{Brown1994,Hillenbrand1997}. The remaining three groups
have intermediate ages, but their exact ages are challenging to
estimate. There are too few very high-mass stars to estimate correctly
the main sequence turn-off mass, and the lower-mass stars have not
evolved significantly. Age estimates for OB1b ranges from 1.7$\pm1.1$ Myr 
\citep{Brown1994} to 8 Myr \citep{Blaauw1964}, see e.g. Table 1 in 
\citet{Caballero2007}. The age of OB1c is reported to be in the range
3-6 Myr\citep{Blaauw1964,Warren1977,Brown1994}, and $\lambda$ Ori 
is comparable to the OB1b and OB1c clusters at approximately 
6 Myr \citep{Dolan2001}. It is thus not clear that there is any
significant age difference between these three groups.

\begin{table*}
\begin{center}
\caption{The 5 regions with massive stars in Orion. Given are the
estimated number of stars above 2$M_{\odot}$, the age for rotating
and non-rotating models, the highest mass possible mass for stars
at these ages, the estimated number of stars that has already
exploded as supernovae and the distance.}
\label{tab:regs}
\begin{tabular}{lcccccc}
\hline\hline
Association & Stars $>2M_{\odot}$   & Age (rotating) & Age (non-rotating) & $M_{up}$ & Stars$>M_{up}$  &  Distance\\ 
\hline
OB1a & 160 &  12 Myr & 10 Myr & 18.5 & 7.3 & 330\\
OB1b & 120 &  5.5 Myr &  4.6 Myr & 45 & 1.3 & 360\\
OB1c & 140 &  5.5 Myr &  4.6 Myr & 45 & 1.5 & 400 \\
OB1d & 60 &   1 Myr & 1 Myr & 120 & 0 & 410\\
$\lambda$ Ori & 60 & 5.5 Myr & 4.6 Myr & 45 & 0.65 & 450\\
\hline
\end{tabular}\\
\end{center}
\end{table*}

\subsection{The currently most massive stars}
Previous studies of the ages of the stellar groups have focused
mainly on less massive stars, due to their much larger numbers.
Also the strong winds from these stars make them more challenging to
analyze, as line blanketing effects have to be taken into account,
which has only recently become possible to do in detail. However,
it is not clear if massive stars arrive at the zero-age main sequence
at exactly the same time as less massive stars.
We therefore compile a list of recent determinations of the properties
of the most massive stars in the Orion region, to analyze the ages
and masses of the stars. The advantage of using these stars is that
unlike lower mass stars, they move significantly in the 
$\log T_{\mathrm{eff}}-\log L$ diagram on a timescale of Myrs. Given a set of 
evolutionary models, one can therefore derive a relatively precise 
evolutionary age even if the observational errors are large. 
On the other hand, the theoretical evolution of these massive stars
is still poorly understood, and assumptions to derive $\log T_{\mathrm{eff}}$ and
$\log L$ from observations introduce relatively large errors.
For the Orion OB1 associations, we include the stars listed in
\citet{Brown1994}, whereas \citet{Dolan2001} was used for $\lambda$ Ori.

Orion hosts 7 O stars altogether. Their properties were analyzed
in previous studies, using photometric data interpreted
using local thermodynamic equilibrium (LTE) models \citep{Kurucz1992}
that were not corrected for line-blanketing. The results provided
by such analysis are, however, very unreliable for massive stars.
We therefore derive new properties based on spectroscopy rather than
photometry, and using line-blanketed non-LTE models rather than LTE:
We use the recent catalogue of \citet{Maiz2004} to identify the spectral 
types of the O stars. We then use the line-blanketed models of 
\citet{Martins2005} to estimate their effective temperatures 
(using the observational scale), luminosities and surface gravities.
We increased the sample by adding the two very bright B stars 
($\epsilon$ Ori A and $\kappa$ Ori) from \citet{Searle2008}.
For the Orion Nebula Cluster a more detailed study of the five most
massive Trapezium stars was performed by \citet{Simon-diaz2006}, and
we use their results for these five stars.

\begin{figure*}
\resizebox{\hsize}{!}{\includegraphics[angle=0]{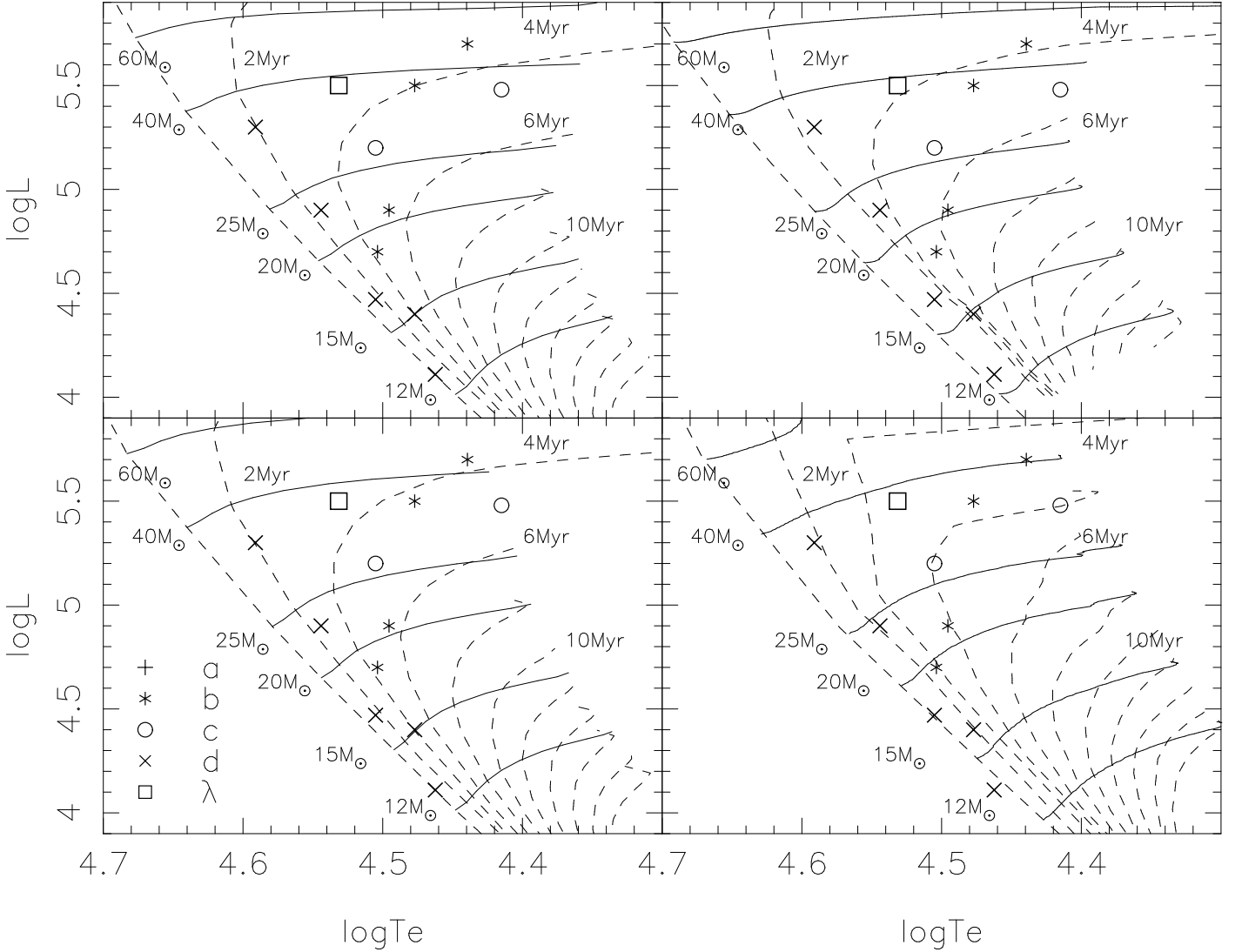}}
\caption{The 7 O stars and 2 of the bright B stars in Orion, compared to
stellar main sequence tracks (solid lines) and isochrones (dashed lines),
from four different stellar evolution models. The tracks correspond to
stellar models of 12,15,20,25,40 and 60 $M_{\odot}$. The isochrones
correspond to ages between 0 and 20 Myr with a distance of 2 Myr.
The stellar models are from \citet{Meynet1997} (upper left),
\citet{Limongi2006} (upper right), \citet{Schaller1992} (lower left)
and \citet{Meynet2005} (lower right). 
}
\label{fig:ostars}
\end{figure*}

\begin{table*}
\begin{center}
\caption{The stars shown in figure \ref{fig:ostars} and their derived
properties, using rotating stellar tracks. Numbers from the Henry Draper
catalogue, commonly used names, spectral types are given, together with
their spectral mass, and the initial and current masses assuming rotating
stellar models.}
\label{tab:ostars}
\begin{tabular}{lccccccc}
\hline\hline
Association & HD & Name & Spectral type & Mass (Spec) & Mass (current) & Mass (Initial) & Age\\ 
\hline
OB1b & 36486 &  $\delta$ Ori A & O9.5II & 17.1 & 21.1 & 21.4 & 5.6\\
OB1b & 37468 &  $\sigma$ Ori A & O9.5V & 15.9 & 19.5 & 19.6 & 3.8\\
OB1b & 37742 &  $\zeta$ Ori A & O9.7Ib & 20.3 & 30.9 & 34.3 & 5.5\\
OB1b & 37128 &  $\epsilon$ Ori A & B0Ia& 45.5 & 34.6 & 40.8 & 5.7\\
OB1c & 37043 &  $\iota$ Ori A & O9III & 19.7 & 26.2 & 27.3 & 5.2\\
OB1c & 38771 &  $\kappa$ Ori & B0.5Ia & 27.3 & 28.0 & 31.8 & 6.2\\
OB1d & 37022 &  $\theta^1$ Ori C & O7Vp & 44.8 & 34.7 & 35.5 & 1.8\\
OB1d & 37041 &  $\theta^2$ Ori A & O9V & 27.5 & 24.0 & 24.1 & 2.2\\
OB1d & 37020 &  $\theta^1$ Ori A & B0.5V & 12.8 & 16.1 & 16.2 & 2.4\\
OB1d & 37023 &  $\theta^1$ Ori D & B0.5V & 18.4 & 17.7 & 17.7 & 0.6\\
OB1d & 37042 &  $\theta^2$ Ori B &B0.5V & 9.5 & 13.4 & 13.4 & 0.0\\
$\lambda$Ori & 36861 &  $\lambda$ Ori A & O8III & 38.8 & 34.0 & 37.0 & 4.2\\
\hline
\end{tabular}
\end{center}
\end{table*}

\begin{figure}
\resizebox{\hsize}{!}{\includegraphics[angle=0]{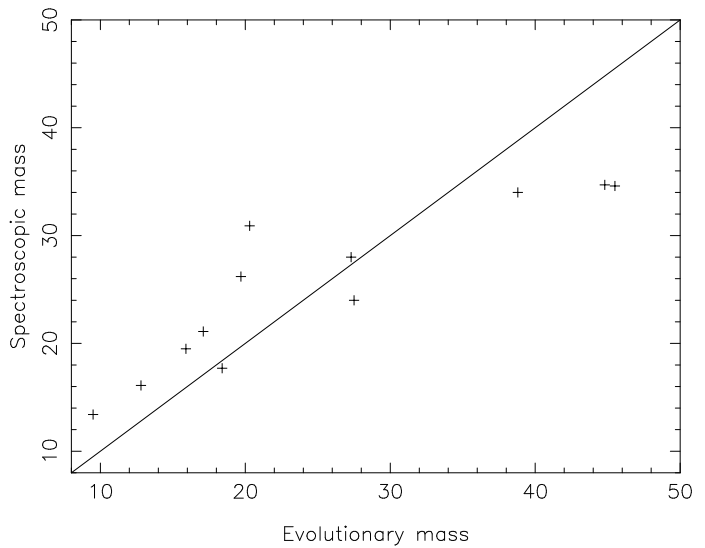}}
\caption{The evolutionary masses of the stars from table \ref{tab:ostars}
versus their spectroscopic masses.
}
\label{fig:masses}
\end{figure}

In figure \ref{fig:ostars} the stars are shown in an $\log T_{\mathrm{eff}}-\log L$
plot, where they are compared to four different sets of stellar evolutionary 
tracks and isochrones, with (upper panels and lower left panel) and without (lower right panel) including
the effects of rotation. Stellar models including rotation are taken from
\citet{Meynet2005} and the  models  without inclusion of rotation from \citet{Meynet1997,Schaller1992,Limongi2006}.
From this plot the masses and ages of the individual stars
can be derived. The masses implied for the stars do not vary significantly
between the models, but a systematic shift in ages between the non-rotating
and the rotating models is apparent. For the
stars in groups OB1b,c and $\lambda$ Ori, the average stellar age is
0.8 Myr higher for models including rotation than for the ones without,
whereas the five stars in group OB1d are on average $\sim0.7$ Myr younger for
the models with rotation. There are no significant differences between the
results obtained from the three different non-rotating models. 
The results for the models including rotation are given in table 
\ref{tab:ostars}. 

For many decades O-star research has been subject to a severe mass 
discrepancy \citep[e.g.][]{Herrero1992}, where spectroscopic masses
(derived from log g) and evolutionary masses (derived from the 
luminosity, and the mass-luminosity relation) were highly discrepant, in 
some cases by more than 50\%. Here we re-investigate this issue for 
Orion's massive star population (see figure \ref{fig:masses}). 
It is comforting to notice that there no longer appears 
to be any significant mass-discrepancy.
 
It is clear that properties of the massive stars in the groups OB1b-d and
$\lambda$ Ori are consistent with ages of a few Myr, whereas there are
no very massive stars in OB1a, in agreement with a higher age of
$\sim$10 Myr. The 4-5 Myr ages of the stars in OB1bc and $\lambda$ Ori fall
approximately in the middle of the age estimates derived from the less 
massive stars, with no evidence for the much lower age of 1.7 Myr
for OB1b found by \citet{Brown1994}. The most massive star in OB1d
is found to be the youngest. The stars in OB1d are all consistent with
an age below 2 Myr.

In table \ref{tab:regs} we summarize the assumptions on the different stellar groups
as we use them in the following. We emphasize that there are
considerable uncertainties on the numbers given in table \ref{tab:regs},
which should be evident from the discussion above and from figure \ref{fig:ostars}. 
As the numbers are gathered from a large
number of sources with varying assumptions and methodology, and many
of these lack reliable estimates of the uncertainties, 
we have not included error estimates
in the table. However, in the following analysis, we do 
estimate the sensitivity of our results to our assumptions.

\subsection{The population of B stars}  %%%%%%%%%%
We also compile a list of stars with masses
between 8 and 20 $M_{\odot}$ for each of the 5 subgroups 
 \citep[from][]{Brown1994,Dolan2001,Hernandez2005},
and assign the assumed average age per group.
We note that the use of \citet{Kurucz1992} atmosphere models in these
earlier studies is not quite appropriate for these relatively massive stars,
so that the uncertainty on these estimates increases significantly.
However, as the output of kinetic energy, matter, and UV radiation
from these B stars is relatively small, compared to the more massive
O-type stars, this is not important for our study.
As the study of \citet{Brown1994} does not give the masses of stars,
we calculate these from:
\begin{equation}
M=\frac{gL}{4G\pi \sigma_{\mathrm{sb}} T_{\mathrm{eff}}^4},
\end{equation}
where $g$, $T_{\mathrm{eff}}$ and $L$ are their given surface gravities, effective
temperatures and luminosities per star, and $G$ and $\sigma_{\mathrm{sb}}$
are the gravitational and the Stefan-Boltzmann constants. 
%The study of \citet{Brown1994} includes all the relevant stars in the 
%study of \citet{Hernandez2005}.  %%% ??? what is that supposed to say?
For the stars present in both the \citet{Brown1994} and \citet{Hernandez2005}
study we use the former to identify the membership of the 
stellar associations, while we consider the masses
given by \citet{Hernandez2005} to be more precise.
Only two stars differ in masses by more than 30\% between the two
catalogues: HD 35439 and HD 37756, which have masses of 34.3 $M_{\odot}$
and 8.8 $M_{\odot}$ from \citet{Brown1994} and 11.3 $M_{\odot}$ and
13.8 $M_{\odot}$ in \citet{Hernandez2005}. For two stars that are only
present in the catalogue of \citet{Brown1994}, we find unrealistically
high masses given their spectral types. 
These are HD 41335 and HD 37061 for which masses of
42.2 $M_{\odot}$ and 50.7 $M_{\odot}$ are found, with spectral
types B2Vne+ and BIV. A number of the stars given in table 
\ref{tab:ostars}, as well
as HD 35439 mentioned above, also have unrealistically high masses
in that study; we ascribe this to the very high
observational uncertainties, and assign both stars a mass of
15 $M_{\odot}$ in our sample. The final list of stars in the 5
regions is given in table \ref{tab:bstars}.  

\begin{table}
\begin{center}
\caption{The observed stars with masses $>8 M_{\odot}$ and which are
not listed in table \ref{tab:ostars}.}
\label{tab:bstars}
\begin{tabular}{lrrrc}
\hline\hline
Association & HD & HIC\footnotemark & Mass & Mass Ref\footnotemark\\ 
\hline
OB1a & 35007 & 25028 & 8.2 & H\\
OB1a & 34748 & 24847 & 8.3 & H\\
OB1a & 35912 & 25582 & 8.5 & H\\
OB1a & 35762 & 25493 & 8.6 & H\\
OB1a & 36351 & 25861 & 9.0 & H\\
OB1a & 35575 & 25368 & 9.0 & H\\
OB1a & 36741 & 26098 & 9.3 & H\\
OB1a & 35777 & 25480 & 9.4 & H\\
OB1a & 35299 & 25223 & 9.9 & H\\
OB1a & 36166 & 25751 & 10.0 & H\\
OB1a & 37490 & 26594 & 11.1 & B\\
OB1a & 35439 & 25302 & 11.3 & H\\
OB1a & 35149 & 25142 & 11.4 & H\\
OB1a & 35411 & 25281 & 11.9 & B\\
OB1a & 35715 & 25473 & 13.1 & H\\
OB1a & 35039 & 25044 & 13.3 & H\\
OB1b & 36827 & 26120 & 8.1 & H\\
OB1b & 36779 & 26106 & 8.5 & H\\
OB1b & 37674 & 26683 & 9.7 & H\\
OB1b & 37744 & 26713 & 9.7 & H\\
OB1b & 37776 & 26742 & 10.0 & H\\
OB1b & 37903 & 26816 & 10.1 & H\\
OB1b & 36695 & 26063 & 12.5 & H\\
OB1b & 37479 & 0 & 12.7 & B\\
OB1b & 36591 & 25980 & 13.6 & H\\
OB1b & 37756 & 26736 & 13.8 & H\\
OB1c & 37040 & 26257 & 8.3 & H\\
OB1c & 36629 & 26000 & 8.4 & B\\
OB1c & 39291 & 27658 & 8.4 & B\\
OB1c & 37209 & 26345 & 8.6 & B\\
OB1c & 37334 & 26442 & 8.9 & H\\
OB1c & 38051 & 26908 & 9.2 & H\\
OB1c & 36959 & 26197 & 9.4 & B\\
OB1c & 39777 & 27929 & 9.6 & H\\
OB1c & 37303 & 26427 & 11.0 & H\\
OB1c & 37018 & 26237 & 11.4 & B\\
OB1c & 35337 & 25202 & 11.6 & B\\
OB1c & 37481 & 26535 & 11.6 & B\\
OB1c & 37356 & 26477 & 11.8 & H\\
OB1c & 33328 & 23972 & 12.7 & B\\
OB1c & 37017 & 26233 & 14.6 & H\\
OB1c & 0 & 0 & 15.0 & B\\
OB1c & 36960 & 26199 & 16.4 & B\\
OB1c & 36512 & 25923 & 16.8 & B\\
OB1c & 41335 & 28744 & 15.0 (42.2) & B\\
OB1d & 36982 & 0 & 8.6 & B\\
OB1d & 37061 & 26258 & 15.0 (50.7) & B\\
$\lambda$ Ori & 37232 & 0 & 9.3 & D\\
$\lambda$ Ori & 34989 & 0 & 11.9 & D\\
$\lambda$ Ori & 36822 & 0 & 17.9 & D\\
\hline
\end{tabular}
\end{center}
\tiny{
$^{1}$HIPPARCOS catalogue \citep{Perryman1997}\\
$^{2}$B:\citet{Brown1994};
 H:\citet{Hillenbrand1997}; D:\citet{Dolan2001}.}
\end{table}

\subsection{The total population of massive stars}
In addition to the observed stars, a number of massive stars are expected
to have formerly existed in the Orion region, and exploded as supernovae
in the last 10 Myr. Knowing the age of an individual region and
the number of stars in a given (lower) mass range, one can
calculate the expected number of higher-mass stars using a distribution function for
initial masses (IMF). In table \ref{tab:regs} we list the
expected number of exploded stars for each of the groups, assuming
a Salpeter IMF \citep{Salpeter1955}.

To test consistency of our inferred stellar content of the groups with observations,
we compare the source lists to the values listed in table \ref{tab:regs}.
We add up the number of individual stars above 8 $M_{\odot}$ in
table \ref{tab:ostars}, table \ref{tab:bstars} and the expected
number of exploded stars from table \ref{tab:regs}: a total of
62 stars is obtained, while 70 are expected from applying the Salpeter 
mass function to the numbers of stars above 2$M_{\odot}$ in
table \ref{tab:regs} (excluding the 11 stars that are expected to
have exploded). We note that the lower mass limits of the
star counts in \citet{Brown1994} are relatively high ($4-7 M_{\odot}$) 
and therefore possible errors due to deviations from the Salpeter
law below $8 M_{\odot}$ are small, The extrapolation of the star counts from 
\citet{Hillenbrand1997,Dolan2001} are more uncertain as the lower mass 
limits in these studies were 1 $M_{\odot}$ and 2.5 $M_{\odot}$.
The upper mass limit depends on the assumed ages of the associations
and stellar evolution models. However, the expected number of massive 
stars is not sensitive to this limit, due to the relatively small fraction
of stars at the massive end of the IMF.
 
The only group where the number of stars
in our list is significantly different from the IMF-expected value
is $\lambda$ Ori, where 4 stars above 8$M_{\odot}$ are observed,
whereas 7.8 are expected. We note that the mass estimates in \citep{Dolan2001} are imprecise for the massive
stars due to inappropriate atmosphere models used, and that several stars are estimated just below the 8$M_{\odot}$ limit.
Another source of bias arises from us using the number of observed OB stars within $5\deg$ to
estimate the richness of the group: a significant fraction of these could
be unrelated to the group, inappropriately scaling up the group richness.

As our list is compiled from various sources, applying different
selection criteria and analyses, it is not appropriate to use it
to estimate the mass distribution function. Nevertheless, we checked 
if our assumed Salpeter initial mass function is compatible with 
our stellar data: We
sort the observed sources into 4 mass bins, and compare to expectations from
a Salpeter and a Scalo IMF. For simplicity the Scalo IMF has been normalized
to have the same normalization at $8 M_{\odot}$ as the Salpeter IMF,
instead of deriving the normalization from the star counts of the
individual associations. The results are shown in table \ref{tab:IMF}. 
Only in the $20-30M_{\odot}$ bin the difference is significant, 
and within uncertainties we consider both 
a Salpeter IMF and a Scalo IMF adequate to represent the overall data.

\begin{table}
\begin{center}
\caption{The number of observed stars in 4 mass ranges, compared
to the expectations according to the \citet{Salpeter1955} and the
\citet{Scalo1986} mass functions.}
\label{tab:IMF}
\begin{tabular}{lccc}
\hline\hline
Mass range & Observed & Salpeter & Scalo\\ 
\hline
$>30M_{\odot}$ & 5 & 4.8 & 2.8\\
$20-30M_{\odot}$& 3 & 7.2 & 4.7\\
$15-20M_{\odot}$& 8 & 10.7 & 8.0\\
$8-15M_{\odot}$& 46 & 47.7 & 42.5\\
\hline
\end{tabular}
\end{center}
\end{table}

\section{Outputs from the massive stars} %%%%%%%%%%%%%%%%%%%
We investigate the ejection of matter, of $^{26}$Al
and $^{60}$Fe, and the UV emission from the stellar groups
in Orion, using the population synthesis method developed by
\citet{Voss-popsyn}. Due to the proximity of Orion,
the populations of stars are relatively well-known, as described
above. We discuss three approaches to calculate
the outputs from the stars in star-forming regions, comparing the
results. In the first approach, 
the Orion stellar population is described by three parameters: the total
number of stars, the average stellar age, and the age spread. 
The second (refined) approach models each of the 5 subgroups separately
with these three parameters. In the third method, we directly
use the observed massive stars with their parameters, together with
estimates of those that have already exploded
as supernovae.

\begin{figure}
\resizebox{\hsize}{!}{\includegraphics[angle=0]{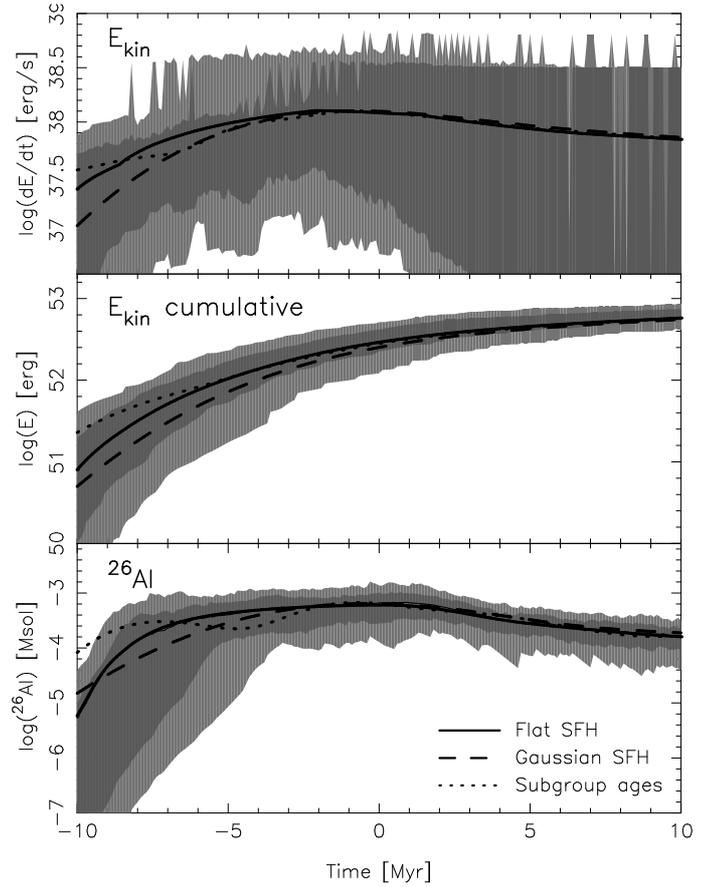}}
\caption{The Orion region modelled as a single cluster (model I) with a flat
and a Gaussian star formation history, respectively, and compared to a model where the age of each subgroup is used (model II). 
From top to bottom are shown the time profiles of the 
differential and cumulative energy ejection and the amount of $^{26}$Al
present in the surrounding ISM. The dark and light grey shaded areas
correspond to 1$\sigma$ and 2$\sigma$ statistical variations of the flat star formation history model, determined from random sampling of the mass function.
}
\label{fig:onecluster}
\end{figure}

In our population synthesis, stellar-evolution tracks are evaluated/interpolated to find the
mass loss and kinetic energy from the stellar winds as a function of time. The supernova
contribution is added as stellar evolution terminates, assuming a canonical ejection energy of $10^{51}$ erg. 
The UV emission versus time is found from matching stellar atmosphere models with
the stellar parameters at a given time.
 In addition to calculating the cumulative stellar outputs for the entire population, 
 we estimate the statistical deviations, caused by the
random sampling of the IMF. Discussions of the shape of the distributions
caused by this can be found in \citet{Cervino2006,Voss-popsyn,Gounelle2009}. 
We compare two ways: An analytical formula 
 developed by \citet{Cervino2006}, and Monte Carlo sampling.
Our method was found consistent both with results from the {\tt Starburst99} code 
\citep{Leitherer1999,Vazquez2005}, and with the results of a similar but different population synthesis implementation by \citet{Cervino2000}.

In \citet{Voss-popsyn} different stellar-model inputs were analyzed and compared. 
In the following, we use three different stellar models, in order to 
represent the possible spread from theoretical predictions.
\begin{itemize}
\item {\tt geneva05}: The stellar-evolution models of 
\citet{Meynet2005,Palacios2005} including the effects of rotation, 
together with the supernova yields of \citet{Limongi2006}.
\item {\tt geneva97}: The stellar-evolution models of
\citet{Maeder1994,Meynet1997} without inclusion of stellar rotation, with {\it enhanced} mass loss, together with
the supernova yields of \citet{Woosley1995} extracted from core sizes, similar to
the method of \citet{Cervino2000}.
\item {\tt LC06}: The stellar-evolution models and
supernova yields of \citet{Limongi2006}.
\end{itemize}
For all three models we use stellar-wind velocities according to {\tt wind08} \citep{Lamers1995,Niedzielski2002}
and the {\tt atmosMS} \citep{Kurucz1992,Martins2005,Smith2005}
stellar atmosphere models.

\begin{figure}
\resizebox{\hsize}{!}{\includegraphics[angle=0]{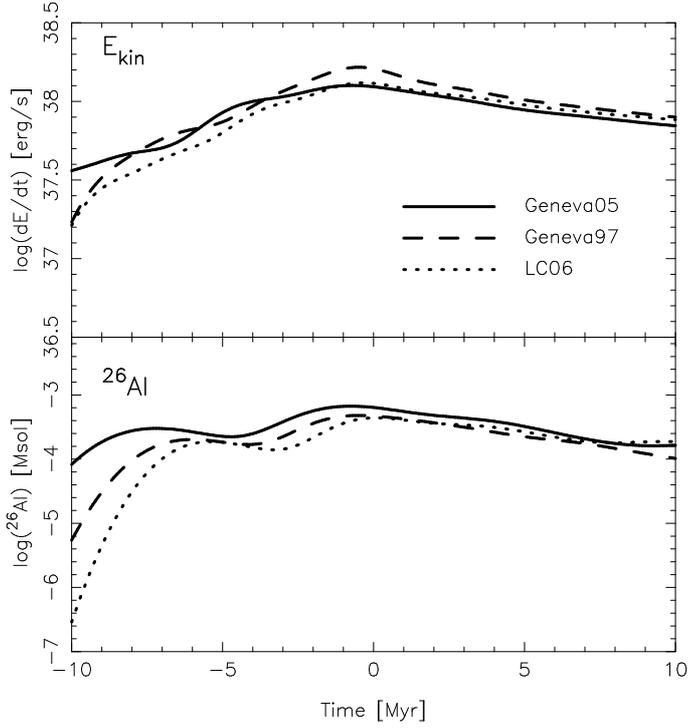}}
\caption{The comparison of the time profiles of the kinetic energy ejection
and the $^{26}$Al present in the ISM for 3 different sets of stellar evolution
models. The individual ages of the subgroups (model II) were used for the calculation.
}
\label{fig:fivecluster}
\end{figure}

%\subsection{Models of Orions stellar population} %%%%%%%%%%%%%%%%%%%%%
\label{sec:models}
We compare three different models for Orion's stellar population:
\begin{itemize}
\item[I.]
{\it Orion as one cluster:}
This corresponds
to how we would have to model more distant star-forming regions, where less
information is available on the individual groups. Given the estimated
numbers of stars in each group from table \ref{tab:regs}, the whole
region is expected to have 81 stars above 8~$M_{\odot}$.
Results from a flat star-formation rate over the last 12 Myr are
compared to a model in which the star-formation rate is Gaussian with
a peak 6 Myr ago and with dispersion of 3 Myr, where we truncate the Gaussian at 2 $\sigma$.
\item[II.]
{\it Orion as five groups:}
Here we treat the 5 (4) separate groups individually.
For each of the groups we use the parameters from table \ref{tab:regs},
and assume a Gaussian star-formation rate with a dispersion of 1 Myr, again
truncated at 2$\sigma$.
\item[III.]
{\it Using the observed stars:}
In this model we use the observed stellar parameters directly,
together with estimates of the stars that have already exploded as
supernovae. We use the derived stellar masses directly (for the stars in
table \ref{tab:ostars} we use the masses derived from the rotating
stellar evolution models), but assign each star the average age of the
association. The most massive stars are generated randomly from 
IMF extrapolation, as above.
\end{itemize}

\begin{figure}
\resizebox{\hsize}{!}{\includegraphics[angle=0]{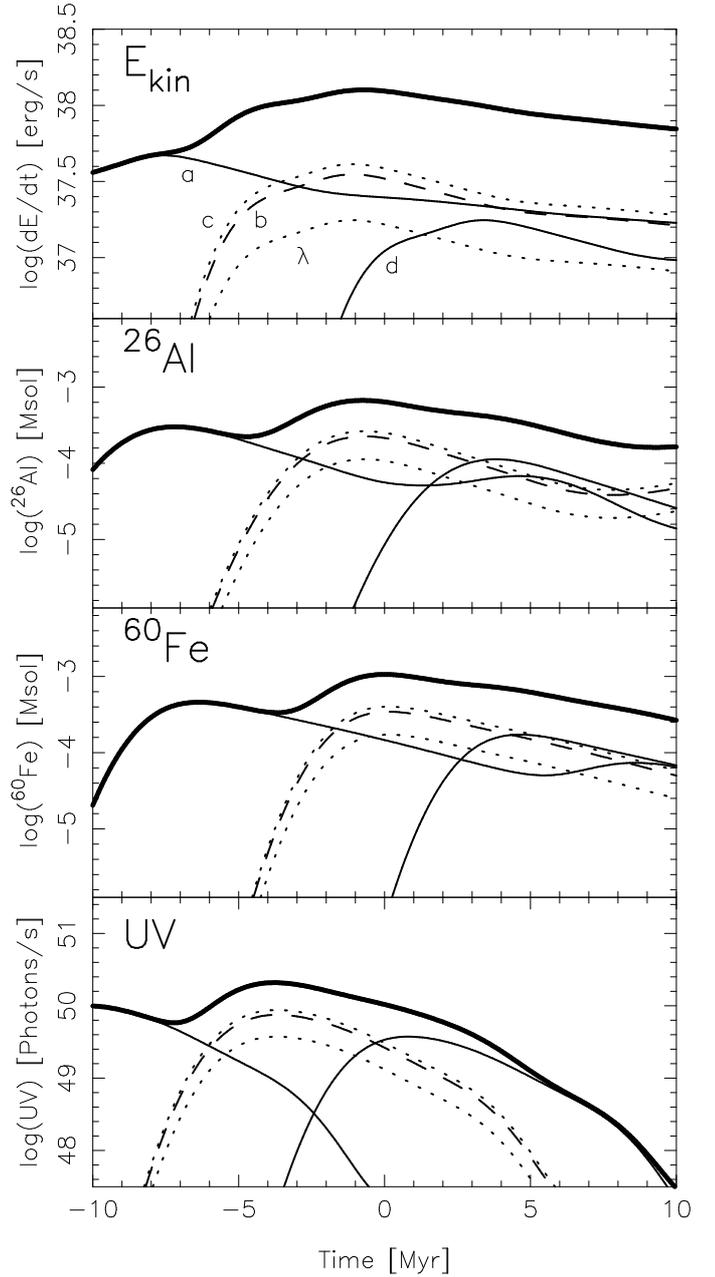}}
\caption{The time profiles of the differential kinetic energy ejection, the
$^{26}$Al and $^{60}$Fe present in the ISM, and the emission of ionizing photons,
from model II.
The contributions from the individual subgroups are shown.
}
\label{fig:subgroups}
\end{figure}

\begin{figure}
\resizebox{\hsize}{!}{\includegraphics[angle=0]{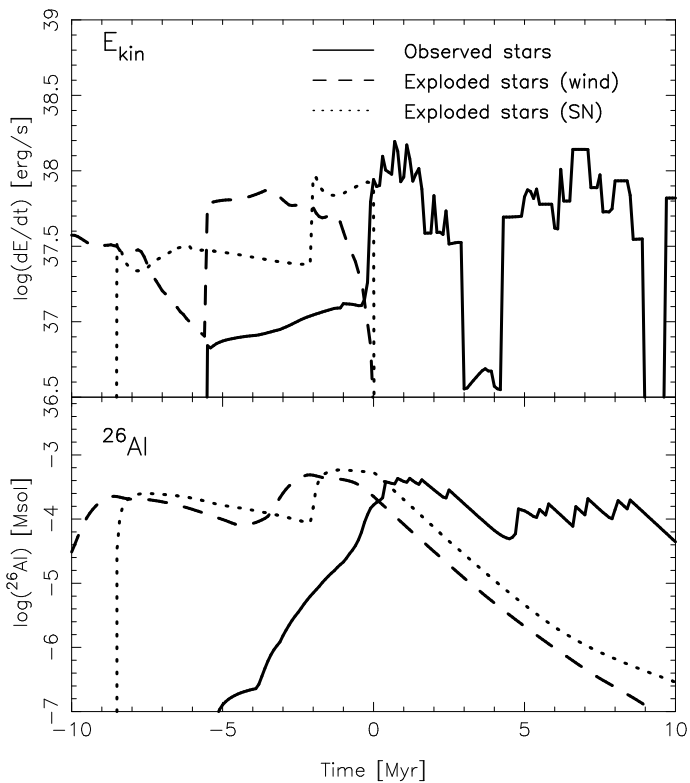}}
\caption{The kinetic energy ejection and $^{26}$Al present in the ISM using
model III.
The figure compares the contribution from currently observed stars
with the inferred contributions from already exploded stars, 
where the latter is divided into
the wind contribution and the supernova contribution.
}
\label{fig:stars}
\end{figure}

\section{Results} %%%%%%%%%%%%%%%%%%%%%%%%%%%%%%%%%%%%
When comparing models of star-forming regions to observations, it is
important to understand how our incomplete understanding of the regions
may affect such comparison. Often, the star formation history is
poorly constrained, but may be important considering the relatively short timescales of interest of
$\sim10$ Myr. In figure \ref{fig:onecluster} we show the
time profiles of kinetic energy ejection from the stellar winds and supernova
explosions and of the interstellar mass of $^{26}$Al. Lines show the three different
star formation histories described in section \ref{sec:models}: 
{\it flat} (model I), {\it Gaussian} (model I) and the {\it separate ages for each subgroup}
(model II). The results of
all three models are surprisingly similar: Values for the current and future times 
 are the same within $\lesssim$10\%. Some differences appear
in the past values, increasing towards the time of formation of
subgroup OB1d (12 Myr ago). We conclude that the
properties investigated in this paper are not sensitive to the exact
star formation history for regions with ages above 5-6 Myr, and they
cannot be used to constrain earlier star formation, accordingly.
Shaded areas in figure 
\ref{fig:onecluster} show the 1$\sigma$ (dark grey) and 2$\sigma$
(light grey) statistical variations. These are derived through random sampling of the mass function \citep[see e.g.][]{Cervino2006},
and are large because both the kinetic energy of the winds and the ejection
of $^{26}$Al strongly depend on the ZAMS mass of the stars.
It is clear that these variations are larger than the uncertainties in
the star formation history. We note that the lines indicate the 
\textit{average} values, and that these probability distributions are strongly
asymmetric for small numbers of stars \citep[see][]{Voss-popsyn}.

In figure \ref{fig:fivecluster} we compare the results of three different
stellar evolution models and supernova yields, that are considered representative
of the spread in theoretical predictions. For all three models the
subgroups were modelled individually (model II).
A spread in current values
of $\sim20-30$\% can be seen, yet much smaller than the statistical
variation. The main differences at early times are between the
stellar evolution models including rotation and the ones without, with both more energy
and $^{26}$Al being ejected from the stars in their wind phases than from their supernovae. This difference is mainly
caused by two effects: the somewhat higher ages of the sub-groups inferred
by the stellar models including rotation (e.g. the 2 Myr higher age of subgroup 
OB1a) and the enhanced wind ejection caused by stellar rotation.

The current output is dominated by the OB1b,c and Ori $\lambda$ subgroups.
At their current age ($\sim4-6$) they still contain very massive stars, but
the most massive stars have exploded relatively recently. OB1a is old
enough to not have any stars above $\sim20 M_{\odot}$ left and is therefore
mainly contributing in the current epoch with supernova output, whereas OB1d is so young that
most stars have not developed strong Wolf-Rayet winds yet and no supernovae has
exploded, so overall contributions are small. 
In figure \ref{fig:subgroups} we show the time profiles of
the emission of kinetic energy and ionizing photons, and the amount
of $^{26}$Al and $^{60}$Fe present in the surrounding ISM, with the
relative contributions from the individual subgroups. 
Some differences
between the behaviour of the different outputs can be noted. For example,
the UV radiation from subgroup OB1a has become totally insignificant,
as this is linked to the most massive stars, whereas the energy and
isotope ejection from supernovae still plays a role. In contrast, subgroup OB1d currently emits a high fraction of the total ionizing
UV radiation, and some kinetic energy and $^{26}$Al, but no $^{60}$Fe,
which is only ejected by supernovae.

The results show clearly that for star-forming regions with $\lesssim100$
massive ($>8M_{\odot}$) stars, the random sampling of the initial 
mass function limits the physical interpretation of
observations. 
For the stars still present today, the actual masses of
the observed stars were used (model III). 
The solid line in Fig. \ref{fig:stars} represents their outputs 
and appears jagged, 
from actual statistical sampling. This is in contrast
to the already exploded stars, where the \textit{average} output 
is inferred based on the IMF. 
In this case spikes caused by supernova explosions are smeared out due
to our lack of knowledge of the actual sampling.  
Clearly exploded stars dominate the
past history of the cluster, whereas the unexploded stars dominate
the future. 
The figure also shows that the current state of the
system is mainly determined by the \textit{exploded} stars. While
the observations of stars are important for understanding the
stellar population of star-forming regions, they can not be used
to reduce the effects of the random sampling of the IMF significantly.

The choice of IMF may have some impact on the results. In
\citet{Voss-popsyn} the contribution of the different parts of
the mass function to the time profiles of $E_{kin}$, UV radiation
and $^{26}$Al and $^{60}$Fe were discussed (see their Fig. 13).
The Scalo mass function would decrease the number of stars above
80 $M_{\odot}$ by a factor of $\sim2$ and the in the $40-80 M_{\odot}$
range by $\sim40$\%. This will significantly delay and flatten the
peaks of the time profiles of the individual groups. Our results for groups OB1b,c and $\lambda$ Ori will be affected.
The results for these groups are currently dominated by the stars in
the $30-50 M_{\odot}$ range, and the values predicted by our models
should therefore be lowered by 30-40\%. 
OB1a is old enough to be in the regime dominated by lower
mass stars (15-30 $M_{\odot}$ range), where the number of stars does not
differ much between the two mass functions. The total content of massive
stars in OB1d is known and evaluated without any IMF consideration.

\section{Comparison with observations}
We compare our model predictions against three observables which characterize feedback in the Orion region: the kinetic energy as manifested in ISM excavation, 
the ionization resulting from UV output, and the radioactive materials $\gamma$-ray luminosity.

The envelopes of the massive stars are ejected through stellar winds
and supernova explosions at typical velocities of a few 1000 km s$^{-1}$
\citep{Woosley1995}, and this energy  
creates large cavities around OB associations. The flows of
supernova ejecta inside cavities can be very complex \citep{MacLow2005},
and the propagation might be dominated by turbulent diffusion from magnetic
field irregularities caused by the stellar winds and supernovae
\citep{Parizot2004,Balsara2005}.
The Eridanus cavity is a typical example of a cavity in the ISM,
created by the cumulative and sustained action of massive stars in the Orion OB1 association.
This interpretation is consistent with the age determinations of the OB1
subgroups and the cavity found by \citet{Brown1995},
who also estimated the energy required for creating the
Orion-Eridanus bubble to be approximately $1.9\times10^{52}$ erg.
Our population synthesis model yields a total
of $1.8^{+1.5}_{-0.4}\times10^{52}$ erg with $\sim40$\% coming from OB1a and
the rest from OB1b,c. OB1d has not yet broken out of the surrounding
medium and $\lambda$ Ori is not connected to the Eridanus bubble.
Our population synthesis results are therefore in agreement
with these observations.

The total flux of the free-free radio continuum emission
can be used as an observable reflecting the total Lyman continuum
luminosity of a region. As Orion covers a very large area
of the sky ($\sim600$ Deg$^2$) only a part of it has been measured by
radio telescopes. Observations of the Greater Orion Nebula
(M42) covering the most luminous parts of Orion OB1d,
report radio luminosities in the 1-25 GHz band of 300-500 Jy
\citep{Felli1993,vanderWerf1989}. This translates to a
emission of hydrogen ionizing photons of $5-8\times10^{48}$ ph
s$^{-1}$ \citep{Condon1992}. This
is significantly smaller than our estimate of $\sim3\times10^{49}$ ph
s$^{-1}$ on average. 
However, the  UV output is strongly dependent on the most massive star
in a cluster, making it highly sensitive to small-number statistics.
Indeed the number of observed massive stars in OB1d is significantly smaller than that from
a population synthesis view, and thus the UV radiation is statistically very uncertain and almost unconstrained,
with a 1-sigma confidence interval of 3$\times10^{47}$-4$\times10^{49}$ ph s$^{-1}$.
The absence of stars more massive than 45 $M_{\odot}$ in
OB1d indicates that the UV radiation should be well below the population-synthesis predicted average
found by integrating over the entire mass function. Indeed an integration
over the expected output from the observed stars yield an ionizing
UV output of $10^{49}$ ph s$^{-1}$. With a leakage of 25\%-50\% of the ionizing photons, similar to what has been inferred
in the Carina region \citep{Smith2007} this number would agree well with
the radio continuum observations. Therefore, we do not consider this a significant discrepancy between predicted and observrd ionizing energy.

The COMPTEL $\gamma$-ray telescope has mapped the all sky distribution 
of the $^{26}$Al decay line at 1.809 MeV emission over 9 years of
observations. The results for the Orion region are presented in
\citet{Diehl2002}. 
Depending on the spatial model, the emission from the Orion region is found at a confidence level of $7-9\sigma$,
and a total flux of $2.8-3.7\times 10^{-5}$ ph cm$^{-2}$ s$^{-1}$ is found.
This corresponds to a mass of $\sim4-5\times10^{-4}$ $M_{\odot}$ of $^{26}$Al at a distance of 400 pc, in good agreement with the results
shown in figure \ref{fig:onecluster}. Calculating the emission separately
for the 5 groups and taking into account their individual distances, we
get an expected flux of $4.5^{+2.1}_{-2.0}\times10^{-5}$ ph s$^{-1}$ from
the OB1 association, in good agreement with the observations. 

A map of the observed signal, although limited by the total signal weakness, shows $^{26}$Al emission in the
Orion region, with a main peak consistent with the position of 
Orion OB1, and extended emission towards lower latitudes,
suggestively aligned with the direction of the Orion-Eridanus bubble. 
Nearly all the flux is coming from the OB1b,c groups which are producing
equally strong signals. A modest
addition of $\sim3\times10^{-6}$ ph s$^{-1}$ is expected to come
from $\lambda$ Ori, which was not included in the observational analysis.

\section{Summary and discussion} %%%%%%%%%%%%%%%%%%%%
We analyzed the population of massive stars in the nearby star-forming
Orion region, including the four OB1 subgroups (a-d) and the
$\lambda$ Ori group. We analyzed the stellar contents of the
individual groups, providing updated lists of the stars more massive
than 8 $M_{\odot}$. Ages of the individual groups were constrained
based on comparison between the updated properties of the most massive
stars and stellar isochrones.
Based on these results, we performed a study of the ejection of kinetic
energy and radioactive elements from the young massive stars in Orion.
We showed that the current state of the region only depends modestly
on the properties of the model, such as the star formation history and
the stellar evolution models. Main uncertainties are due to the
unknown population of very massive stars that exploded over the past
10 Myr. 

The population synthesis results were compared to the
energy needed to form the Eridanus superbubble, the emission of
hydrogen ionizing photons, and the intensity of
the 1.809 MeV line from the decay of $^{26}$Al, showing good agreement
between our model estimates and the observations. The $^{26}$Al observations
provide a valuable tracer of the population of (now not any more observable) stars and
thus of the cumulative action of massive star groups, and 
of the kinematics of the outflows from the massive stars.

Our current understanding of stellar evolution and supernova
models is far from complete. Different models often
rely on similar assumptions. Showing consistency between
models and observations is important, as it supports confidence 
that the most important effects are accounted for in models.
We have employed different models for characterizing the Orion region's stellar population,
and for the stellar-evolution inputs to population synthesis. The results show that
the observed properties of the Orion region are consistent
with these models. Differences among models
are smaller than the statistical effects caused by the
relatively small number of massive stars. 

Some recent UV studies
\citep{Bouret2003,Fullerton2006} have called
for a more fundamental mass-loss rate reduction, invoking clumping factors
up to $\sim$100, much higher than the currently favoured values of
$\sim 5$, and mass-loss rate reductions of order 10. However, other studies
cast doubt on these conclusions based on theoretical studies of
"macro-clumping" \citep{Oskinova2007,Sundqvist2010} and emission in
the extreme UV band \citep{Waldron2010}.
As we find good agreement between our population synthesis and the
observations of Orion, this could either 
suggest that our mass-loss rates are realistic (and the very large 
clumping factors exaggerated), or alternatively that
some unknown process is also missing in the stellar models.
However, we note that the wind and supernova contributions to the
interstellar $^{26}$Al have not been disentangled observationally,
and models with weak winds to some degree compensate for the low 
$^{26}$Al wind yields by having larger core masses and therefore
producing higher supernova yields \citep[see discussion in][]{Limongi2006},
and the production of $^{26}$Al in high-clumping models have not 
yet been explored. On the other hand, we emphasize that our models are 
in \textit{simultaneous} 
agreement with both the kinematic and the radioactive tracers, which would
be hard to achieve with models involving very large clumping factors.

The Galaxy contains hundreds of regions of massive star-formation. 
It is important to extend our approach to other regions in order to overcome 
the issue of small-number statistics, and to further test our models. 
Unfortunately, many such regions are either significantly less well studied than Orion, due to larger distances and
obscurance from the foreground, or they only contain modest numbers of
high-mass stars. Recent studies of the relatively small, but nearby
Scorpius-Centaurus region \citep{Diehl2010} and the more distant but
very massive Cygnus region \citep{Martin2009} have been reported, and show overall agreement between
the observations and models, both regarding the energetics of the
regions and the $^{26}$Al signal. 
A further candidate target is
the Carina region \citep{Smith2006}, hosting a large population of 
very young and very massive stars. Due to the small age, the supernova contribution to
the $^{26}$Al signal in this region is expected to be low. A
comparison between Orion and Carina $^{26}$Al signal could therefore potentially 
constrain the relative wind and supernova contributions, similar to what has been done in the Cygnus region \citep{Martin2009}.
  
Radioactive tracers are a valuable addition to the
arsenal of probes of star formation in the Milky Way. $\gamma$-ray
observations have the potential to yield information that is complementary
to observations at other wavelengths. They are emitted on a timescale of
Myr after star formation, and with similar decay timescales, they
trace the cumulative action of very young stars in the Milky Way. As the
$\gamma$-rays are unaffected by extinction, the observations of
$^{26}$Al and $^{60}$Fe have the potential to give a complete view
of the star formation in the Milky Way, unaffected by the obscuring
effects of the molecular clouds.
Using nucleosynthesis ejecta, 
we can expand the studies of past activity from stellar groups 
substantially. 
Instrument sensitivities of current $\gamma$-ray telescopes limit such studies to the brightest, hence most-nearby regions. 
A next generation of instruments \citep{Greiner2009}
could reach hundreds of massive-star regions, thus significantly extending such studies.
As discussed in \citet{Voss-popsyn} the correlation
between $^{26}$Al and $^{60}$Fe can potentially eliminate
much of the uncertainty due to small-number statistics. 
The lifetimes of the radioactive tracers are long enough that they
can be carried to significant distances from the massive stars that
produced them. The COMPTEL observations of Orion \citep{Diehl2002}
provide a hint of such displacement. On the other hand the lifetimes
are short enough that the radioactive elements are not uniformly
mixed into the ISM. They are therefore valuable tracers of the mixing
processes in the vicinity of star forming regions.

\begin{acknowledgements}
This research was supported by the DFG cluster of excellence 'Origin and
Structure of the Universe' (http://www.universe-cluster.de). 
\end{acknowledgements}

\end{document}